\documentstyle[11pt,aaspp4]{article} 

\newcommand{\etal}{{\it et al.\/}}

\begin{document}

\title{Caltech Faint Galaxy Redshift Survey XIV:
Galaxy Morphology in the HDF (North) and its Flanking Fields to $z=1.2$\altaffilmark{1}}

\author{Sidney van den Bergh\altaffilmark{2},
Judith G. Cohen\altaffilmark{3}, 
David W. Hogg\altaffilmark{4,5,6} and  Roger Blandford\altaffilmark{4}}

\altaffiltext{1}{Based in part on observations obtained at the
	W.M. Keck Observatory, which is operated jointly by the California 
	Institute of Technology and the University of California}
\altaffiltext{2}{Dominion Astrophysical Observatory, National Research Council,
    5071 West Saanich Road, Victoria, British Columbia, V9E 2E7,
    Canada:sidney.vandenbergh@nrc.ca}
\altaffiltext{3}{Palomar Observatory, Mail Stop 105-24,
	California Institute of Technology, Pasadena, CA \, 91125
        jlc@astro.caltech.edu}
\altaffiltext{4}{Theoretical Astrophysics, California Institute of Technology,
	Mail Stop 130-33, Pasadena, CA \, 91125 rdb@tapir.caltech.edu}
\altaffiltext{5}{Current Address:  Institute for Advanced Study, Olden Lane, Princeton, NJ \, 08540 hogg@ias.edu}
\altaffiltext{6}{Hubble Fellow}

\begin{abstract}
Morphological classifications are reported for Hubble Space
Telescope (HST) images of 241 galaxies in the Hubble Deep
Field (HDF) and its Flanking Fields (FF) with measured redshifts in 
the interval $0.25 < z < 1.2$, drawn from a magnitude-limited redshift survey
to $R = 24.0$. 
The galaxies are divided into 
three groups with redshifts in the intervals [0.25,0.6], [0.6,0.8],
[0.8,1.2]. R$_{606}$ images from the first group and I$_{814}$ images 
from the 
second and third groups are compared with B-band images of nearby galaxies. 
All classifications were therefore made at approximately the same rest
wavelength. Selection biases are discussed.

We corroborate and extend the results of earlier investigations by observing that:
\begin{itemize}
\item Most intermediate and late-type galaxies with $z\gtrsim0.5$ have morphologies that 
are dramatically different from those of local galaxies and cannot 
be shoehorned into the Hubble ``tuning fork'' classification scheme.
\item Grand-design spirals appear to be rare or absent for $z\gtrsim0.3$.
\item Many Sa and Sb spirals with $z\gtrsim0.6$ do not exhibit well-defined 
spiral arms.  The arms of distant Sc galaxies appear more chaotic
than those of their nearby counterparts.
\item The fraction of all galaxies that are of types Sc and Scd drops
from 23\% at $z \sim 0$ to 5\% for $z > 0.6$.  
\item Barred spirals are extremely rare for $z\gtrsim0.5$.
\item Roughly one in five galaxies with $z\gtrsim0.8$ are compact objects 
that resemble local E, S0 or Sa galaxies.
\item Peculiar galaxies are more common beyond $z=0.3$,
especially among late-type spirals, than they are at $z\sim0$.
\item Merging galaxies, particularly those with three or more components,
also become more common with increasing redshift.
\end{itemize} 

On the basis of these and similar observations, it is inferred 
that the development of pronounced spiral structure was delayed until
$\sim5$~Gyr and that most bulges are probably not formed by 
disintegrating bars.  Major morphological changes were 
still taking place only $\sim5$~Gyr ago even though changes in the integrated light 
of most galaxies were then much slower than they were $\sim10$~Gyr ago. 
\end{abstract}

\keywords{ galaxies: evolution, galaxies: formation, surveys}

\section{Introduction}

The Hubble Space Telescope has, for the first time, allowed us
to observe directly the evolution of galaxy morphology over a
significant fraction of the age of the Universe. Soon after 
the images of the Hubble Deep Field North, HDF, and its Flanking Fields (FF)
became available (Williams \etal\ 1996, Ferguson, Dickinson \& Williams 2000)
it was clear that distant and young field galaxies were 
very different from their local, contemporary counterparts (Abraham \etal\ 1996a,b).
Indeed, half a lifetime of experience in galaxy
classification (van den Bergh 1998) proved to be of only marginal
usefulness in attempts to classify galaxies at intermediate and large
look-back times. In the words of Ames (1997), ``You are living in a land you no 
longer recognize. You don't know the language.'' More specifically, it was concluded that 
``The fraction of interacting and merging objects is seen to be significantly 
higher in the Hubble Deep Field than
it is among nearby galaxies. Barred spirals are essentially absent from the 
deep sample. The fraction of early-type
galaxies in the Hubble Deep Field is similar to the fraction of early-types 
in the Shapley-Ames Catalog, but the
fraction of galaxies resembling archetypal grand-design late-type 
spiral galaxies is dramatically lower in the distant
HDF sample.'' (van den Bergh \etal\ 1996). 
Because no redshifts were available, it was not possible 
to establish a timescale for galaxy evolution. It is the purpose of the present
paper to re-examine these conclusions, using the HDF and FF images and 
the new redshift information
(Cohen \etal\ 2000 and references therein), and to discuss the galaxy evolutionary timescale.

The initial reactions to the HDF images have been largely vindicated and significantly 
developed over the past four years. (See Abraham 1999, 2000 for 
excellent reviews.)
The most complete information so far available on
the morphological evolution of galaxies with $z\lesssim1$
is by Brinchmann \etal\ (1998) who classified
HST images of a complete sample of 341 galaxies, selected from both the
CFRS and LDSS surveys, for
which ground-based redshifts were available. They found
a substantial increase in the fraction of irregular
galaxies from $\sim$9\% at $z \sim$0.4 to $\sim$32\% at $z \sim$0.8
and associated these galaxies with the increase in blue luminosity 
density with redshift.  However,
the reliability of this conclusion is undermined 
by the fact that the standard ``irregular'' galaxy
that Brinchmann \etal\ use to calibrate their
classifications appears to be an Sb pec galaxy
with a central bulge and the kind of under-developed 
spiral structure that is typically seen
in spirals at large redshift. 

Using the same sample, Lilly \etal\ (1998) found that the sizes
of large galaxy disks do not change significantly out to $z \sim$1 ,
but that the rate of star formation was elevated by a
factor 3 at $z\sim$0.7. In a new approach, Abraham \etal\ (1999a) related the evolution of 
galaxy morphology to their internal star formation as monitored by 
their spatially-resolved colors. They confirmed 
that spiral bulges pre-date their disks and followed different 
evolutionary histories.
In a study that also included the HDF South, Abraham \etal\ (1999b) 
confirmed that barred spirals were comparatively rare prior to $z\sim 0.5$.

Ellipticals, by contrast,
appear to have evolved relatively little in density and luminosity since $z\sim1$. 
There is evidence for ongoing and declining star formation although less than 
5\% of their 
stars are estimated to have formed over this time (Schade \etal\ 1999). They seem to have formed
over an extended interval $1\lesssim z\lesssim3$ (Odewahn \etal\ 1996, Driver \etal\ 1998). 
This is consistent with the work of Corbin \etal\ (2000)
who find that 12 out of 111 (11\%) of their NICMOS images (in a data
sample that extends to photometric redshifts as large as $z = 2.7$) are probably
elliptical galaxies.

Turning to irregulars, the HST images of
285 CFRS/LDSS galaxies have been analyzed to derive the 
evolution of the merger fraction out to $z\sim$1 (Le F\'evre \etal\ 2000). 
Up to 20\% of luminous galaxies are found to be
in physical pairs (some of which may be projections)
at $z{\sim}0.8$. A typical L$^\ast$ galaxy was found to have undergone $\sim1-2$ mergers
since $z\sim1$. However, the ``chain'' galaxies, first identified by Cowie, Hu \& Songaila (1995), appear
to be neither edge-on spirals nor merger products (Abraham \etal\ 1999b).

In recent years, there has been a shift away from the traditional, descriptive 
and, inevitably, somewhat subjective morphological approach
towards more quantitative measures of galaxy structure like scale lengths, central concentration,
asymmetry, and so on. This is particularly valuable for connecting 
observations to increasingly sophisticated numerical simulations. However,
galaxies are too variegated to be 
completely described by just a few numbers and we believe that 
a simple morphological approach will continue to be of value, 
even at high redshift.
After all, the durability of the original Hubble (1936) scheme for nearby galaxies is remarkable 
given all that we have learned about them since the 1930s. 

In this paper we continue to use the traditional morphological classification, 
despite its manifest inadequacy at high redshift, for three reasons.
The first is quite modest, to increase our confidence that galaxies at high redshift are
quite different from local galaxies by enlarging 
the sample size. The second is to determine if there are {\it any} counterparts to certain
local galaxy types at high redshift. For example, 
if it is believed on dynamical grounds (see \S 4.1) that grand 
design spirals take a minimum of $\sim10$~Gyr to grow, then the discovery of just one {\it bona fide}
example at $z\sim1$ would be highly 
significant. The third reason is that it is important to determine
the chronology of these changes and, in the 
absence of an adequate high redshift classification, the best 
approach is to compare with the well-understood local 
classification scheme and to discover
when it starts to fail.

In the following section, we describe our procedure and discuss some possible selection biases.
In section 3, we give our empirical conclusions and we conclude with a brief interpretation of their
implications for physical theories of galaxy formation.

\section{Galaxy Classification}

\subsection{Morphology Sample}

Our sample is based on the redshift survey in the region of the HDF
of Cohen \etal\ (2000), which includes 671 objects. 
This survey is 92\% complete with respect
to the photometric catalogs of Hogg \etal\ (2000) to $R \le 24$ in the
HDF itself, and also 92\% complete
to $R \le 23$ in a region 8 arcmin in diameter centered on the HDF.
See Cohen \etal\ (2000) for details of the samples, the number
of stars, galaxies, and high redshift ($z > 1.5$) galaxies, AGNs and QSOs
in the HDF and in the Flanking Fields.
\footnote{The assembly of the sample for the present morphological
study was carried out before the final version
of Cohen \etal\ (2000) was available.  A few galaxies ($\sim10$) whose
redshifts were only determined in the late fall of 1999 that are
included in Cohen \etal\ (2000) are not
included here.}

The sample used for morphology and discussed here
is divided into three redshift intervals, $0.25<z<0.60$, 
$0.60<z<0.80$ and $0.80<z<1.20$. In each interval, the sample contains
all available galaxies in the spectroscopic redshift catalog with
suitable HST images.  Thus the low redshift group contains 49 galaxies of which 43 are drawn 
from the HDF.  Four galaxies are in the PC field of the WFPC2 image and
two others are just outside the boundary adopted for the HDF but
within the coverage of the HST $R$ image.   (There
is unfortunately no $R$ HST image of the rest of the area of the Flanking
Fields.)  Apart from this there is no bias in magnitude, color or location in this 
subsample. The intermediate redshift group includes 70 galaxies, while
the high redshift group includes 120 galaxies.  Note that the 
WFPC2/HST I$_{814}$
images do not cover the entire area of the Flanking Fields, and hence
some galaxies within the area of the redshift survey could not be used.
The fraction of galaxies drawn from the HDF and from the FF is as
would be expected based on the relative areas and the rise in
galaxy counts as a function of $R$ magnitude in this magnitude regime.

\subsection{Biases and selection effects}

From a practical point of view, the morphological classification
of individual galaxies at $z\sim1$ presents three distinct challenges 
beyond similar classifications of nearby systems. The first is that the 
images used for the classification
of nearby galaxies might contain as many as $\sim$10,000 pixels, whereas
those of very distant Hubble Space Telescope (HST) images may 
contain only $\sim$100 pixels and apparent changes in fine scale features may 
simply be a consequence of resolution. 
Furthermore, due to longer  exposure times the quality of the 
HDF images is markedly superior to that of the FF images.
This may lead to some misinterpretation of the morphologies
of the faintest FF galaxies, especially the most compact examples. 
For example, it is often not possible to distinguish Galactic
stars with certainty from compact ellipticals of Hubble types E0 and
E1. (All of the objects in the present sample, including those
that have morphological classification ``star'' or ``E0/star'' are extragalactic, on the basis of their redshifts,
though.) 
%
%
The comparative statements below are restricted to those 
that can be made with 
confidence on the basis of the poorest images. (We note, in passing, that it is the 
large dynamic range of CCD detectors
makes them particularly suitable for use in galaxy classification, van den Bergh 
\& Pierce 1990). 

The second challenge is to correct for band-shifting. Galaxy classification 
is traditionally performed in the B (440nm) band. We find, ({\it cf} also Fig. 3 from
Ferguson \etal\ 2000) that galaxy morphology varies
sufficiently slowly with wavelength that it is adequate to compare R$_{606}$ images from 
the 0.25-0.60 redshift interval (the shift is exact for $z=0.38$)
and I$_{814}$ for $0.60<z<1.20$ (exact for $z=0.85$) with local samples.
In particular, the $z$ dependence of the frequency of barred spirals 
(\S\ref{barred-spirals}) cannot
be attributed to band-shifting, as has sometimes been suggested 
(e.g. Bunker \etal\ 2000, Eskridge \etal\ 2000).
Some small band-shift effects could, however, still be experienced for 
galaxies that have $1.0\lesssim z\lesssim1.2$. In such objects
the core/halo ratio might be depressed, and giant stellar associations
may appear slightly enhanced, relative to similar structures seen at
smaller redshifts.  

The third challenge is that, in a magnitude-limited survey, 
we are comparing more luminous 
objects at high redshift with less luminous objects at low $z$. 
We emphasize that 
our local comparison sample is not the HDF but the more extensive Shapley-Ames catalog.  With respect to the sample of the
Revised Shapley-Ames Catalog, the median $M_B$ (rest frame) for
the HDF sample is about 0.3 mag fainter than that of the RSA, where
the median for the RSA sample is taken from
figure 5 of Sandage and Tammann (1981) and has been adjusted to 
our adopted value of H$_0$.
(Only RSA galaxies with known redshifts as of that date are plotted in this
figure.)  A correction for internal absorption was applied
to the late type spirals in the RSA sample, while none has been applied here.
Hence the two samples are comparable in their luminosity range;
we are in fact seeing quite far below $L^{\ast}$ in the HDF sample.

Relative to the median galaxy in $R$ at $z=0.3$ the absolute magnitudes of the median
galaxies with $z=$0.6, 0.8, 1.2 are $-1.8$, $-2.6$, $-3.6$ smaller
due solely to the larger distance using our adopted cosmology.
\footnote{In this paper, we assume a flat universe with $\Omega_M=0.3,h=0.7$ and age $t_0=14$~Gyr.}
The median (in $z$) galaxies 
in the intermediate and high redshift intervals have 
absolute magnitudes $-1.1$ and $-1.9$ smaller
than the median in the low $z$ interval. It can be argued that about half of this 
luminosity selection is appropriate because it compensates passive stellar evolutionary effects.
However, taking a strictly morphological approach, we have ignored this and simply
make comparative observations about galaxies with similar luminosities
to those at high redshift. Our prime conclusions are quite robust to this choice.
However, luminosity selection is a serious concern when probing the evolution of
secondary characteristics such as disk surface brightness (Simard \etal\ 1999).

\subsection{Sampling}

An additional and related caveat concerns the size of the sample and the sensitivity
to fluctuations resulting from the luminosity selection. A good measure of this is the 
number density of $L^\ast$ galaxies, specifically $\Phi^\ast V_c$, where $V_c$ is the comoving
volume and $L^\ast$, $\Phi^\ast$ are determined locally, 
expected in our three redshift 
intervals ({\it cf} Ferguson \etal\ 2000). This corresponds 
to 8, 10, 20 bright galaxies 
on the HDF and nine times as many galaxies in the FF. However, galaxies are strongly clustered and the 
fluctuations over small comoving volumes are much larger than Poissonian. Furthermore,
the HDF was carefully selected to avoid bright galaxies (Ferguson \etal\ 2000) 
and there is a clear deficit of these for $z\lesssim0.3$ (Cohen \etal\ 2000). In particular the nature of
the galaxian population in the fields studied in the present
investigation might have been affected by the vagaries of the 
density-morphology relation (Dressler 1980) along the line of sight.
This concern dictated our choice of $z=0.25$ as the lower limit of our redshift intervals.

\subsection{Morphological typing}

The morphological classifications reported in this paper were made by SvdB, 
and are on the DDO system of van den Bergh (1960abc). The images were 
supplied to him as 20'' x 20'' thumbnails along with the redshift interval 
and no other information.   The areal scale on the sky per pixel 
was the same for all the images.
All images were inspected, excepting those where there were
serious crowding or edge effects. The classifications for the three redshift intervals, 
along with commentary, are 
presented in Tables 1-3.  The $R$ magnitudes
from Hogg \etal\ (2000), the redshifts from Cohen \etal\ (2000),
and the rest frame $B$ luminosities from Cohen (2000)
are also given.  See Cohen (2000) for the
definition of the rest frame luminosity and how
it is derived. 

The fractions of galaxies by type are collected in Table 4.
(Intermediate types, such as Ir/Merger, are counted as 0.5 Ir and 0.5 Merger.)
In the present paper we have, following van den Bergh \etal\ (1996), 
used the somewhat judgemental term  ``protogalaxy''
to denote objects that resemble the  prototypical galaxy  
H36555\_1249, which is shown in 
        Figure 7. A color image of this object, which is
        shown as Plate 9 of van den Bergh \etal\ (1996), shows
        what appears to be a reddish off-center nuclear bulge 
        that is embedded in a rather chaotic looking disk 
        which contains half a dozen bright blue knots. Objects 
        with this type of morphology were regarded as spirals 
        that are still in the process of being assembled. In
objects described as ``mergers'' the individual components/knots have 
not yet combined to form a single coherent structure.

\section{Galaxy morphology at high and low redshift}

As explained in the introduction, we interpret distant galaxies 
by reference too their local counterparts, as exemplified by the 
Shapley-Ames galaxies (van den Bergh
1960c, Table 4). The differences are striking.

\subsection{Spirals}

The present sample of galaxies shows
an almost complete absence of ``grand design'' spirals, i.e. disk systems with
well-developed long arms of DDO types Sb I, Sbc I and Sc I
({\it cf} van den Bergh \etal\ 1996). Previous
experience (van den Bergh 1989), which was confirmed by radial velocity
observations (Visvanathan \& van den Bergh 1992), showed that such
luminous grand design spirals could have been recognized on images that
contain as few as $\sim$100 silver grains.

Even
less pronounced spiral structure, such as is seen in nearby spirals
of DDO luminosity classes II, III and IV, appears to be rare in distant
HDF + FF galaxies. Furthermore the spiral structure that is observed
in such distant spirals appears to be more chaotic ({\it{i.e}} less regular)
than do Sc and Scd spirals at $z\sim 0$.
The fraction of Sc, Sc/Ir galaxies increases 
from $\sim5$\% at $z\sim$1 to $\sim10$\% 
at $z\sim$0.5 and $\sim$~23\% at $z\sim$0.
Most of the missing Sc, Scd and Sc/Ir galaxies at $z \sim1$ are 
probably masquerading under the categories ``Protogalaxy'',
``Peculiar'' or ``Merger''.
In intermediate and early-type ``spirals'' little (or no) evidence is
actually seen for spiral arms. At $z \sim$0.8 this effect might be partly
(but not entirely) due to the low spatial resolution with which
galaxies at such large redshifts are viewed. However, at $z \sim$0.4 the
almost complete absence of well-developed spiral arms is certainly
real.  This is so because the mean luminosity of the galaxies in the
present sample increases with redshift. Since the
strength of spiral structure increases with luminosity
one would actually have expected the strength of spiral
arms to increase with $z$. In view of the fact that 
little or no spiral structure is actually seen at $z{\gtrsim}0.5$ the 
Sa, Sb and Sc classifications listed in
Tables 2 and 3 are almost entirely based on central concentration
of light in the galaxy images. In other words spiral arm tilt and
resolution could usually not be factored in to the present
classifications of the most distant early-type and intermediate-type
spirals.

\subsection{Barred spirals \label{barred-spirals} }

Another striking feature of the present HDF + FF sample is the almost complete 
absence of barred spirals at high redshift.  The highest redshift galaxy that
almost certainly has a real bar is at $z = 0.321$, 
although two possible barred galaxies in Table~1 have
redshifts $z = 0.457, 0.475$ and one from Table~3
has $z = 0.962$.
\footnote{ Bunker \etal\ (2000) claim that this object (H36483\_1214),
``seems to be a chance alignment of a swath of young stars 
with the approximate axis of the true bar.''}
Taken at face value this result suggest that only $\lesssim1$\% 
of all high redshift galaxies
are barred spirals, compared to 21\% to 34\% among nearby spirals 
(van den Bergh 1998, p.43). 

\subsection{Ellipticals}

Roughly one fifth of the galaxies with $z \sim1$
are compact, and have morphologies similar to those of nearby E, S0 and
Sa galaxies.  The data
in Table 4 show no statistically significant variation with redshift in
the fraction of field galaxies with classification types
E, E/S0 and S0. 

It is interesting to note evidence from the morphological classification
for at least one density enhancement along the line of sight to the HDF which
contains a clump of nine E0 -- E3 galaxies with $<z> = 0.679$.  This is
one of the most prominent peaks in the redshift distribution of galaxies
in the region of the HDF, and was noticed as such in 
the initial redshift survey analysis of Cohen \etal\ (1996).

\subsection{Peculiar-merging galaxies}

Distant galaxies are far more likely to be classified as peculiar.
Van den Bergh (1960c) found only 
31 out of 540 (5.7\%) of all nearby spirals of types
Sa, Sb and Sc in the Shapley-Ames catalog to be peculiar.  After
excluding edge-on objects (in which it is difficult to detect any
peculiarities), one finds that 17.5 out of 20.5 (85\%) of all spirals
with types Sa--Sab--Sb--Sbc--Sc and redshifts of $0.25 < z < 0.60$ 
are noted as being peculiar in Table~1. Only four out of seven objects of types
Sa--Sab are peculiar, whereas all spirals of types Sb--Sbc--Sc are found
to be peculiar. Perhaps surprisingly, Table 4 shows that the
fraction of peculiar galaxies appears to remain approximately constant
over the range $0.25 < z < 1.2$. The reason for this is, no doubt, that
it is much easier to see peculiarities in the relatively large images
of nearby galaxies at $z \sim0.4$ than it is to notice similar peculiarities
in the much smaller images of very distant galaxies with $z  \sim1$.
 
Similarly, galaxies classified as
probable mergers, and which account for only $\sim$1\% of the galaxies at 
$z\sim$0, represent $\sim$6\% of those with $0.25<z<0.80$ and for 14\% 
at $z\sim1$. 
In particular, there is a rapid increase with redshift in the fraction of triple
and multiply interacting galaxies. In fact, some small regions of the
Hubble Deep Field could be described as debris fields filled with
fragments that might eventually merge into one or more major galaxies. 
\section{Discussion}
Butcher \& Oemler (1978) first established that cluster galaxies
evolve with time. The observations presented here (and in earlier studies
cited above), firmly establish that the morphology of field galaxies also
evolves over time. The challenge now is to use these empirical observations
to infer a physical description of galaxy formation and evolution.
(This approach is quite complementary to deductive methods that use 
``semi-analytic'' extensions to gravitational instability theory 
to make quantitative comparison with measured properties of galaxies.)
The most natural way to address this problem is to take the 
galaxies we see around us today and to ask, in a statistical fashion,
what are their histories. However, this will almost certainly lead to an incomplete
view because galaxies interact and merge and are otherwise strongly
affected by their environment. In addition many galaxies may become too
dim to be observable locally.

\subsection{Spirals}

The low fraction of Sc and Sc/Ir galaxies at high redshift and 
the almost complete absence 
of luminous, ``grand-design'' galaxies beyond $z\sim0.3$, when the universe had an 
age of $\sim10$~Gyr, suggests that a major fraction of the future Sc galaxies
are still classified as protogalaxies or mergers at $z\sim$1, ($t\sim8$~Gyr).
In other words many Sc galaxies had not yet fully assembled at this time.
This is supported by the observation that most early-type spirals in
the field  already appear to have achieved a more-or-less ``normal''
morphology by $z\sim0.4$, while late-type spirals still look peculiar.
However, it should be emphasized that the individual bits and pieces
from which Sc galaxies were assembled might contain quite old
stars and clusters.

This deduction, in conjunction with the fact that significant
numbers of compact early-type galaxies are observed at high
redshifts, may support the view of Boissier \& Prantzos (2000) that
the time-scale for the formation of low-mass disks is longer than
that for more massive ones. This, in turn, is consistent with the view (Bell
\& de Jong 2000) that it is the surface density of a young galaxy that
drives its rate of star formation. As Boissier \& Prantzos point out,
this interpretation may present some difficulties for the standard picture
of hierarchical cosmology. 

The observation that spiral structure is often poorly developed 
in galaxies with $z > 0.5$ is perhaps not surprising.
The rotational period in the outer parts of a giant
galaxy ranges from $\sim0.3-1$~Gyr
and density wave theory requires several rotational periods for the arms to develop 
fully. Even if a disk forms when the universe is, say, 2~Gyr old, there may be only time for 
$\sim5$ rotational periods before it appears in our high redshift interval. 
This constraint is particularly relevant to the grand design spirals, 
which may not assemble 
as disks till the universe was $\gtrsim5$~Gyr old and then 
take another $\sim5$~Gyr to 
develop their arms. In this connection it is of interest to note 
(Fasano \etal\ 2000) that the
galaxian population of rich clusters changed significantly even more recently,
with spirals being transformed into S0 galaxies between $z \sim0.1$ and $z \sim0.25$. Such a transformation of spirals into S0 galaxies might have 
been produced by ram pressure stripping (Gunn \& Gott 1972) or by 
galaxy harassment and tidal effects (Abadi, Moore \& Bower 1999, Quillis, 
Moore \& Bower 2000).
%

\subsection{Barred spirals}

Our observations suggests that most genuine barred spirals only started to form
$\sim4$~Gyr ago. This may be because galaxies at large
$z$ are still too young to have formed dynamically cool disks
which permit bar-like instabilities to develop. In this
connection it is of interest to note that a number of Sb galaxies at
$z \sim1$ appear to have real bulges. (Good examples are F36254\_1519),
F36481\_1102
and F37088\_1117).
It would be important (but difficult)
to do accurate photometry of these spirals to determine if the cores
in these galaxies are true R$^{1/4}$ bulges 
{\footnote {Falomo \etal\ (1997) have previously established that a compact
galaxy with $z = 0.19$ has an R$^{1/4}$ profile.}}
or if they are just the
brightest inner parts of exponential disks. If the presence of true
bulges at $z \sim1$ were to be confirmed this would militate against the
hypothesis of Raha \etal\ (1991) and of Pfenniger, Martinet
\& Combes (1996) that some bulges might have been formed from bars.

\subsection{Ellipticals}

This study does not shed much light on the evolution of elliptical
galaxies, which appear to have mostly formed by $z\sim1$. The major 
uncertainty is whether the compact galaxies that are classified as E, S0 or
Sa at $z\sim1$ are really bulges or if we are already observing disks
({\it cf}~Marleau \& Simard 1998). This is a crucial test for theories 
of galaxy formation.  However,
this issue is complicated by the observation  (Graham \& Prieto 1999)
that the bulges of many late-type spirals are best described  by
exponential luminosity profiles.

\subsection{Peculiar-merging galaxies}

Among nearby NGC galaxies perhaps only $\sim$0.1\%
are members of multiple interacting-merging systems that resemble
Stefan's Quintet (NGC 7317-19). On the other hand a significant
fraction of the present sample of HDF and FF objects appear to be
members of (or associated with)  multiple merging systems. This
result is consistent with the view (Toomre 1977, Abraham 1999,
Carlberg \etal\ 2000, Le F\'evre \etal\ 2000) that mergers were much more
frequent in the past than they are at the present time. However, debate
continues ({\it eg} Carlberg \etal\ 2000) about the rate at which such mergers
increase with redshift, and regarding the mass range of merging
galaxies.

An important caveat is that some close pairs may consist of
projected superposed images of
unrelated galaxies.  In all four cases where the redshifts of both
of a pair of galaxies SvdB considered to be mergers are known, the
redshifts are discrepant, and the objects do not appear to be
physically associated.  (See the notes to Tables 1, 2 and 3.)
\footnote{In one of the four cases, one
of the redshifts is uncertain.  In the other three cases, Cohen \etal\ (2000)
regard both redshifts as secure.}
This serves again as a warning that, in the absence of measured
redshifts for all component galaxies, only obvious tidal 
distortions should move a ``merger suspect''
into the ``merger'' category.

\subsection{Morphological classification at high redshift}

Van den Bergh \etal\ (1996) pointed out that many of the
images of very distant HDF and FF  galaxies do not fit comfortably
into the Hubble (1936) classification scheme which we now find to be 
only strictly applicable to field galaxies with $z{\lesssim}0.3$. In particular, since spiral
structure was rare (or absent) prior to this time, the DDO luminosity
classification system (van den Bergh 1960abc, Sandage \& Tammann 1981)
is not especially useful. 

Can we devise a galaxy classification scheme that would have been useful to sapient beings
who might have been surveying galaxies in their
neighborhood some 5 Gyr ago?
Unfortunately the search for such a system is made more
difficult by the fact that it is often not clear if adjacent luminous
clumps should be regarded as separate galaxies, or as condensations
within a single larger protogalaxy. In fact such a dichotomy is not
even physically significant if initially separate ancestral bits and
pieces eventually merge into a single galaxy. Possibly a two-parameter
classification system based on central concentration of light
(Morgan 1958, 1959) and asymmetry (Abraham \etal\ 1996b) or
clumpiness, perhaps supplemented by color information, may represent
the best that can be done regarding the classification
of galaxies in the early universe.
In rich nearby clusters only the
central concentration of light appears to be a useful classification
parameter. 
  
In this study we have used redshift data and the FF images to 
substantiate many of the original reactions to the first inspection of the HDF
over four years ago. In addition we have shown that the 
development of spiral structure,
both wound and barred, is delayed until the universe is typically
$\sim$10~Gyr old. Perhaps the most pressing current task is to determine if the 
compact objects observed at $z\sim1$ are disks or spheroids.
Overall, though, we remain impressed by the strong evolution in galaxy morphology.
As Hartley (1953) wrote in {\it{The Go-between}} ``The past is a
foreign country; they do things differently there.''

\acknowledgements   It is a pleasure to thank Peter Stetson
for his help with the re-scaling of the present images, 
Simon Morris for cosmological consultations, and Jerry Sellwood
for information on the formation of galactic bars.
The extragalactic work of JGC, who is the PI of the Caltech Faint
Galaxy Redshift Survey, is not supported by any Federal agency.
RB acknowledges support under NSF grant AST-9900866.

%
%
\begin{deluxetable}{llllll}
\tablenum{1}
\tablewidth{0pt}
\tablecaption{Redshifts and Morphological Classifications of}
\tablecaption{Red Images of Galaxies with $0.25 < z < 0.60$}
\label{tab1}
\tablehead{\colhead{ID\tablenotemark{a}} & \colhead{Redshift\tablenotemark{b}} 
& \colhead{$R$\tablenotemark{c}} & \colhead{Log[$L(B)$]\tablenotemark{d}}
& \colhead{Classification} & \colhead{Comments} \nl
\colhead{ } & \colhead{ } & \colhead{(Mag)} & \colhead{(W)} \nl
}
\startdata
F36427\_1306 & 0.485 & 22.02 & 36.10  & Sab/S0 (edge-on) & Projected on merger remnant \nl
F36446\_1304 & 0.485 &  21.14 & 36.57 & Sb pec + Ir/Pec  &    Knots, but no arms,in disk\tablenotemark{e} \nl
F36454\_1325 &  0.441 & 22.33 & 36.31 &  Pec  &             High SB  \nl
F36458\_1325 &  0.321  & 20.71 & 36.23 & Sc pec   &         Only rudimentary spiral  \nl
    &    &    &    & &              structure in disk \nl
F36563\_1209 & 0.321 &  23.22 & 35.34 & Sb pec  &  Knots,but on arms,in disk  \nl
F36575\_1212 & 0.561 & 22.62 & 36.17 & Ir/Merger \nl
H36413\_1141 & 0.585 & 21.91 & 36.32 & Sab   &  Merging with H36414\_1142\tablenotemark{f} \nl
H36414\_1142 & 0.548 & 23.51 & 35.92 & Merger   &  At least three components\tablenotemark{f} \nl
H36416\_1200 &  0.483 & 25.03 & 35.16 &  S/Ir  \nl
H36419\_1205 & 0.432  & 20.82 & 36.63 & Sb pec &   Faint outer arms surrounding  \nl
    &    &    &  & & bright core with incipient arms \nl
H36429\_1216 &  0.454 &  20.51 & 36.79 & Sc pec  & High SB. Two chaotic arms  \nl
H36439\_1250 &  0.557 & 20.84 & 36.89 & Sc pec  &   Knot plus incipient arm in disk  \nl
H36442\_1247 &  0.555 &  21.40 & 36.64 & S  pec  &  Proto-bulge but no arms (yet?) \nl
H36448\_1200 &  0.457 & 22.85 & 35.89 & S(B?)cd: \nl
H36465\_1203 &  0.454 & 24.32 & 35.44 & Sa pec   &  Asymmetric + tidal debris \nl
H36465\_1151 &  0.503  & 22.00 & 36.42 & E1  \nl
H36470\_1236 &  0.321 &  20.62 & 36.42 & S pec &   E3-like core embedded in chaotic \nl
    &    &    &   & &    spiral(?) envelope \nl
H36472\_1230 &  0.421 & 22.63 & 35.91 &  Sab pec \nl 
H36480\_1309 &  0.476 &  20.43 & 36.92 & Sa:  \nl
H36489\_1245 &  0.512 &  23.48 & 35.76 & E:5 pec  &  Asymmetric core embedded in fuzz \nl
H36493\_1311 &  0.477  & 21.97 & 36.30 &  E1 \nl
H36494\_1316 & 0.271 &  23.63 & 35.16 & Pec \nl
H36496\_1257 &  0.475  & 21.91 & 36.31 & Sa pec &   Asymmetric, multiple nuclei \nl
H36497\_1313 & 0.475 & 21.46 & 36.38 & Sb pec    &    Asymmetric \nl
H36501\_1239 & 0.474 & 20.43 & 36.87 & Pec  &  Has high SB central region  \nl
H36508\_1251 & 0.485 & 23.15 & 35.89 & ?  & Disk containing multiple knots  \nl
H36508\_1255 &  0.321 & 22.27 & 35.83 &  S pec &  Nucleus plus two knots in disk \nl
H36513\_1420 &  0.439  & 23.22 & 35.74 & Sb pec   &  Disk with no spiral arms. Off-center \nl
   &    &    &    & &                   nucleus \nl
H36516\_1220 & 0.401 & 21.45 & 36.32 & Sab pec  \nl
H36517\_1353 &  0.557 & 21.08 & 36.80 & Sb pec/Merger &    Double nucleus  \nl
H36519\_1209 &  0.458 & 22.75 & 35.94 & Pec   &  E1-like bulge embedded in \nl
  &    &    &       & &                    asymmetric envelope \nl
H36519\_1400 &  0.559 & 23.03 & 36.00 & Sb pec (edge-on) & Off-center core \nl
H36526\_1219 & 0.401 & 23.11 & 35.72 & E:4 \nl
H36528\_1404 &  0.498  & 23.45 & 35.82 & Pec/Merger \nl
H36534\_1234 &  0.560  & 22.78 & 36.11 & Sb pec &   Off-center bulge  \nl
H36536\_1417 & 0.517 & 23.36 & 35.82 & Sb pec (edge-on) &  Nucleus + two knots in disk \nl
H36549\_1314 &  0.511 & 23.81 & 35.68 & Pec (edge-on)  &  Disk with nucleus + two knots \nl
H36551\_1311 &  0.321 & 23.58 & 35.33 & Pec   \nl
H36554\_1402 &  0.564 & 23.08 & 35.96 & Sbc pec  &  Off-center nucleus + one knot  \nl
  &    &    &     & &                in disk \nl
H36555\_1359 &  0.559 &  23.74 & 35.73 & Sb (edge-on)  \nl
H36560\_1329 &  0.271 &  23.80 & 35.16 & Ir/Pec  \nl
H36566\_1245 &  0.518  & 20.06 & 37.11 & Sb pec  &  Bright E2-like bulge without \nl
  &    &    &     & &       spiral arms. See Fig. 8 \nl
H36569\_1258 &  0.520 &  23.84 & 35.76 & Sa  \nl
H36571\_1225 &  0.561 & 22.36 & 36.27 &  Sc pec  & Has one incipient spiral arm \nl
 &    &    &   & &                 + three knots. See Fig. 9  \nl
H36572\_1259 &  0.475 & 21.07 & 36.61 & SB? pec  &   Bar or two knots \nl
H36580\_1300 &  0.320 & 22.04 & 35.89 & Merger  &     Two nuclei  \nl
H36587\_1252 &  0.321 & 20.99 & 36.27 & SBbc pec &  Probable bar in asymmetric two- \nl
  &    &    &   & &       armed spiral. See Fig. 10  \nl
H36594\_1221 &  0.472 &  23.53 & 35.67 &  S? + Ir:\tablenotemark{g} \nl
H37005\_1234 &  0.563 & 21.43 & 36.74 & S0/Star \nl
\enddata
\tablenotetext{a}{Names are Habcde\_fghi for objects in the HDF, where the object's
J2000 coordinates are 12 ab cd.e +62 fg hi.  The initial letter is
``F'' for objects in the flanking fields.}
\tablenotetext{b}{Redshifts are from Cohen \etal\ (2000).}
\tablenotetext{c}{$R$ magnitudes are from Hogg \etal\ (2000).}
\tablenotetext{d}{Rest frame $B$ luminosities are from Cohen (2000).  
M$_B = -21.0$ (rest frame) $\equiv$ log[$L(B)$ (W)] = 36.9.}
\tablenotetext{e}{The Sb pec is the object for which a redshift exists.
The redshift of the fainter second galaxy is unknown.}
\tablenotetext{f}{The redshifts of this pair are not consistent
with a merger, but the redshift of H36414\_1142 is uncertain.}
\tablenotetext{g}{This is a close pair of faint galaxies separated
by about 1 arcsec.  It is not clear which of them (or perhaps both of them)
was included in the spectroscopic observations.}
%
%
\end{deluxetable}

%
%
\begin{deluxetable}{llllll}
\tablenum{2}
\tablewidth{0pt}
\tablecaption{Redshifts and Morphological Classifications of}
\tablecaption{Infrared Images of Galaxies with $0.60 < z < 0.80$}
\label{tab2}
\tablehead{\colhead{ID\tablenotemark{a}} & \colhead{Redshift\tablenotemark{b}} 
& \colhead{$R$\tablenotemark{c}} & \colhead{Log[$L(B)$]\tablenotemark{d}}
& \colhead{Classification} & \colhead{Comments} \nl
\colhead{ } & \colhead{ } & \colhead{(Mag)} & \colhead{(W)} \nl
}
\startdata
F36194\_1428 &  0.798  &  22.60 &     36.61 &  Ir?   &   Multiple nuclei  \nl
F36244\_1454 & 0.628  &  20.34 &     37.34 &   E2 \nl
F36243\_1525 & 0.682  & 22.78 &     36.48 &   E:3 pec   &     Asymmetrical \nl
F36247\_1510 & 0.641 & 20.41 &     37.26 &    Sa \nl
F36249\_1252 & 0.631 & 22.76 &     36.22 &  Ir (edge-on) \nl
F36250\_1341 & 0.654 &  24.32 &     35.73 & Star/E0  \nl
F36254\_1519 &  0.642  &  21.82 & 36.60 &   Sb pec &  Slightly asymmetrical  \nl
F36270\_1509 & 0.794 &   21.60 &     37.36 &  Sa  \nl
F36275\_1418 & 0.751  &  22.37 &     36.54 &   S (edge-on)  \nl
F36284\_1037 & 0.760 & 22.76 &     36.39 &  E0   \nl
F36287\_1357 & 0.639  & 23.05 &     36.10 & Sb pec?  \nl
F36290\_1346 & 0.693 & 23.02 &     36.25 & E:2   &        Compact  \nl
F36297\_1324 & 0.758   & 23.25 &  36.23 &  Pec/Merger?   & Member of a compact group \nl
F36297\_1329 &  0.748   & 23.03 &  36.44 &  Pec &  Member of a compact group \nl
F36299\_1403 & 0.793  &   21.97 &     36.91 & Merger? \nl
F36334\_1432 & 0.748  &  23.08 & 36.31 &  S: pec  &    In compact group  \nl
F36340\_1054 & 0.762  & 21.55 &   36.87 &  Ir or protogalaxy &  In a group \nl
F36362\_1319 & 0.680  & 22.20 &     36.44 & Lumpy protogalaxy? \nl
F36370\_1159 &  0.779  &  21.56 &     36.84 &  Proto Sc? \nl
F36379\_0922 &  0.767 & 21.43 &     37.08 & Sb: pec &  Has two ``nuclei'',not a bar! \nl
F36384\_1312 & 0.635  &  22.27 &     36.38 &   Sb pec   \nl
F36390\_1006 & 0.635  &  21.22 &     36.81 & ?    &     Has double core  \nl
F36405\_1003 & 0.749   &  22.39 &     36.58 & Sc: \nl
F36415\_0902 & 0.713   &  22.31 &     36.59 & ? \nl
F36427\_1503 & 0.698  &  23.17 &     36.18 & Pec \nl
F36454\_1523 &  0.683  &  22.06 &     36.57 &  S: \nl
F36481\_1102 & 0.650 &  22.58 &     36.40 & Sb \nl
F36481\_1002 & 0.682  & 21.92 &     36.60 & Ir/Pec  \nl
F36499\_1058 & 0.684 &  22.63 &     36.36 & Sa:  \nl
F36575\_1210 & 0.665 &   21.10 &     36.98 &  E:0 + Sb: & 2nd gal. is F36575\_1211\tablenotemark{e} \nl
F36580\_1137 & 0.681 & 23.00 &     36.17 &  E:1  \nl
F36588\_1434 & 0.678 & 20.85 &     37.14 &  S pec  \nl
F36598\_1449 & 0.762  &  21.62 &     37.04 &  S + Sb  &          Merger \nl
F37015\_1129 & 0.779  & 21.45 &     37.07 & Merger  & Has three nuclei. In group \nl
F37017\_1144 & 0.744  &  22.10 &     36.71 &  Sa:   &     Multiple nuclei? \nl
F37020\_1517 & 0.744 & 23.56 &     36.24 & Sab: \nl
F37036\_1353 &  0.745   & 21.63 &     36.78 & Sb pec  \nl
F37058\_1317 & 0.753  & 21.95 &     36.74 & Sc: \nl
F37061\_1332 & 0.753  &  21.85 &     36.83 &  Sbc pec  &  Multiple nuclei?  \nl
F37069\_1208 & 0.693  & 24.13 &     35.78 &  E:0  \nl
F37072\_1214 & 0.655  &  22.19 &     36.62 & Sb  \nl
F37074\_1356 &  0.752   & 23.65 &     36.19 &  Sa    \nl
F37080\_1246 & 0.654 & 21.80 &     36.64 &  Sb \nl
F37083\_1320 & 0.785  &  22.86 &     36.44 & Pec     &   Tadpole-like  \nl
F37088\_1117 & 0.639  &  23.05 &     36.18 &  Sb pec &   Asymmetrical  \nl
F37088\_1214 & 0.788  &  23.90 &     36.05 &  Sa  \nl
F37105\_1141 &  0.789  &  21.20 &     37.10 & Sb pec + S  \nl
F37107\_1431 &  0.677  &  22.39 & 36.61 & E1  \nl
F37108\_1059 & 0.747   & 24.25 &     36.05 & ? + ?   &         Low SB  \nl
F37113\_1545 & 0.692  &  22.43 &     36.42 & dIr?      &     Low SB \nl
F37115\_1042 & 0.778  & 21.97 &     36.74 & Sbc pec \nl
F37163\_1432 & 0.635    & 22.50 &     36.30 & Sb pec  &    Asymmetrical \nl
F37192\_1143 & 0.784   &  22.81 &     36.47 & Sb pec  \nl
F37213\_1120 &  0.656   & 22.22 &     36.47 &  Sc pec \nl
H36389\_1219 & 0.609  &  22.14 &     36.43 &  Pec  & High SB. Complex core = merger? \nl
H36436\_1218 & 0.752 &   22.56 &     36.46 & E1 \nl
H36438\_1142 & 0.765  &   21.26 &     37.32 &  E1  &     In group \nl
H36459\_1201 & 0.679   & 23.88 &     35.85 & Pec  &  High SB. Has two cores \nl
H36470\_1213 & 0.677  &  24.63 &     35.51 &   Pec  &  Core + asymmetrical fuzz  \nl
H36471\_1414 &  0.609   &  23.92 &     35.82 & E5 \nl
H36487\_1318 & 0.753   &  22.87 &  36.40 & Sc pec  &  Very lumpy structure. In cluster \nl
H36494\_1406 & 0.752  &  21.95 &     36.83 &  E3 \nl
H36498\_1242 & 0.751 & 24.38 &     35.64 &  dIr (edge-on) &   In cluster  \nl
H36502\_1245 &  0.680  & 21.74 &     36.86 & E3     &       In cluster \nl
H36538\_1254 & 0.642  & 20.95 &     37.01 & S pec     &    High SB \nl
H36555\_1245 &  0.790   & 23.08 &     36.79 &  S  \nl
H36586\_1221 & 0.682 &  23.40 &     36.03 & E2 \nl
F37222\_1124 & 0.786 & 22.33 &     36.71 & Sa \nl
\enddata
\tablenotetext{a}{Names are Habcde\_fghi for objects in the HDF, 
where the object's
J2000 coordinates are 12 ab cd.e +62 fg hi.  The initial letter is
``F'' for objects in the flanking fields.}
\tablenotetext{b}{Redshifts are from Cohen \etal\ (2000).}
\tablenotetext{c}{$R$ magnitudes are from Hogg \etal\ (2000).}
\tablenotetext{d}{Rest frame $B$ luminosities are from Cohen (2000).  
M$_B = -21.0$ (rest frame) $\equiv$ log[$L(B)$ (W)] = 36.9.}
\tablenotetext{e}{The redshifts of the two galaxies are both known and
not consistent with a merger.}
%
%
\end{deluxetable}

%
%
\begin{deluxetable}{llllll}
\tablenum{3}
\tablewidth{0pt}
\tablecaption{Redshifts and Morphological Classifications of}
\tablecaption{Infrared Images of Galaxies with $0.80 < z < 1.20$}
\label{tab3}
\tablehead{\colhead{ID\tablenotemark{a}} & \colhead{Redshift\tablenotemark{b}} 
& \colhead{$R$\tablenotemark{c}} & \colhead{Log[$L(B)$]\tablenotemark{d}}
& \colhead{Classification} & \colhead{Comments} \nl
\colhead{ } & \colhead{ } & \colhead{(Mag)} & \colhead{(W)} \nl
}
\startdata
F36175\_1402 & 0.818 &  21.73 &     36.93   &  Merger \nl
F36176\_1408 &  0.848  & 22.55 &     36.71 & ? \nl
F36271\_1001 &  0.842 &  21.53 &     36.90 & S pec/Merger? \nl
F36285\_0951 &  1.016  &  23.07 &     36.65 & Sc \nl
F36287\_1023 &  0.936  & 22.11 &     36.91 & Merger/Protogalaxy &   Prototype. See Fig.1 \nl
F36287\_1239 &  0.880  & 22.11 &     36.91 & ?  &    Image too small to classify \nl
F36296\_1420 &  1.055   & 24.00 &     36.06 & Star \nl
F36313\_1113 &  1.013  & 22.09 &     37.07 & Ir/Merger? & Has double nucleus and \nl
   &   &   &  & &    a close companion   \nl
F36332\_1235 &  1.140  & 23.11 &     36.66 &  Ir/Merger?  \nl
F36336\_1005 &  1.015   & 22.15 &     37.12 & Ir/Merger     &   Prototype. See Fig.2 \nl
F36335\_1319 & 0.845  &  21.78 &     37.08 & Sc:      &    High SB spiral with lumpy disk \nl
F36340\_1045 & 1.011  & 23.25 &     36.56 & ?   &  Has nucleus and high SB, \nl
 &  &  &     & &                               and possibly a tail \nl
F36341\_1305 &  0.847 & 24.24 &   36.13 &  Sb + ?  &        Has tidal companion of low SB \nl
F36343\_1312 &  0.845 &  23.15 &     36.52 & S/Ir? \nl
F36364\_1237 &  0.961  & 22.94 &     36.81 & S  &     Spiral with small bulge \nl
 &  &  &    & &                  embedded in lumpy ring \nl
F36367\_1347 &  0.960  & 20.32 &     37.53 & Star\tablenotemark{e} \nl
F36367\_1213 &  0.846  & 20.86 &     37.44 & S IV:   &  Looks like nearby low- \nl
 &  &  &  & &  luminosity spiral \nl
F36369\_1346 &   0.846  &  21.20 & 37.14 & Sb pec  &  Has two nuclei. May be a \nl
 &  &  &    & &                           merger with tidal arms \nl
F36377\_1149 &  0.838  & 22.72 &     36.62 & Ir (edge-on)  &     Has clumpy structure \nl
 & & &      & &                        and no nucleus \nl
F36381\_1116 &  1.018 &  22.20 &     36.95 &   Late-type  &        Spectrum is that of  \nl
 &   &   & & Ir      &              the combined light of  \nl
 &  &    & &      Ir    &                all three components  \nl
F36382\_1150 &  0.842 & 22.74 &     36.75 & Ir   &   High SB irregular \nl
F36388\_1118 &  0.934  & 22.53 &     36.62 & Merger  &           Has three components + \nl
 & & &    & &                        possible tidal arm \nl
F36388\_1257 &  1.127  & 22.28 &     37.22 & Ir    &       Has high SB  \nl
F36399\_1250 &  0.848  & 22.59 &     37.02 & Proto-spiral?  &    High SB object with nucleus \nl
 & & &  & &                and fuzzy tail?  \nl
F36399\_1029 &  0.935 &  22.59 &     37.02 &  ?   &                  Nucleated object with \nl
 & & &  & &                                 possible tidal tail  \nl
F36408\_1054 &  0.875 & 22.61 &     36.69 &  ?   &    Compact,almost stellar  \nl
F36411\_1314 &  1.017  &  23.08 &     36.74 & E1/Sa  &            Nucleated objecte embedded \nl
 & & &   & &                          in asymmetric fuzz  \nl
F36417\_0943 &  0.845 &  22.51 &     37.00 & Sab  \nl
F36420\_1321 &  0.846 &  23.95 &     36.15 & Sa:   &       Very compact \nl
F36425\_1121 &  0.845  & 23.03 &     36.46 & Sa pec    &     Asymmetric \nl
F36435\_1532 &  0.847  & 22.82 &     36.60 & Sa pec? \nl
F36447\_1455 &  0.845  & 22.97 &     36.50 & ?  &    CCD bleeding from bright
     nucleus?  \nl
F36459\_1101 &  0.936 &  22.77 &     37.09 & E + S   &   Interacting pair. Spectrum \nl
 &  &  &  & &      probably of combined light  \nl
F36462\_1527 &  0.851  & 22.11 &     36.94 & Ir?  &             Might also be galaxy disrupted \nl
 &  &  &  & &                 by encounter with nearby E(?)  \nl
F36468\_1540 &  0.912  & 22.23 &     36.99 & Merger?  &      Possible double core suggests \nl
 & & & & &                               this may be merger remnant \nl
F36469\_0906 &  0.905  & 23.84 &     36.38 & Sa:  \nl
F36472\_1628 &  0.873 & 21.69 &     37.46 &  E2 \nl
F36477\_1045 &  1.187  & 23.43 &     37.14 & Sa t?  \nl
F36482\_1507 &  0.890  & 22.38 &     37.20 & E4:  \nl
F36486\_1141 &  0.962  & 22.21 &     37.47 & Sb t  &     Has close compact companion  \nl
F36502\_1127 &  0.954 &  22.88 &     36.62 &  dIr? & Companion to F36518\_1125\tablenotemark{f} \nl
F36518\_1125 &  0.919 & 21.62 &     37.13 & dIr \nl
F36522\_1537 &  0.936  & 22.74 &     37.26 & E0/Star \nl
F36524\_0919 &  0.954  & 22.81 &     37.31 & E3 \nl
F36529\_1508 &  0.942 &  22.84 &     37.15 & E0 \nl
F36532\_1116 &  0.942 &  22.08 &     37.15 & Merger? &           Complex internal structure. \nl
 & & &     & &             Has close compnions \nl
F36539\_1606 &  0.851  & 22.84 &     36.57 & Pec     &       Near edge of image \nl
F36541\_1514 &  0.849  & 22.84 &     36.51 & Pec/Ir \nl
F36548\_1557 &  1.132 & 22.51 &     37.41 & Sab (edge-on) \nl
F36577\_1454 &  0.849  & 22.49 &     36.96 & Sa  &   Near edge of image  \nl
F36583\_1214 &  1.020  & 23.79 &     36.40 & Sa pec   &   Slightly asymmetric \nl
F36589\_1208 &  0.853 &  22.32 &     37.00 & Sb pec    &  One-armed spiral with \nl
 & & &  & &                   off-center nucleus \nl
F36595\_1153 & 1.021 &   22.54 &     36.88 & Sb: t?  &         Two companions included \nl
 & & & & &                     in spectrum  \nl
F37003\_1616 &  0.913  & 22.33 &     36.94 & Merger?  &          Edge-on Ir or merger of \nl
 & & & & &                         three components \nl
F37007\_1106 &  0.801  & 22.94 &     36.41 & Pec &           Contains one ``hot'' pixel \nl
F37016\_1225 &  0.973  &  23.95 & 36.15 & Ir/Merger? \nl
F37016\_1146 &   0.884   & 25.30 &     35.90 & Amorphous  &  Might be high SB Ir \nl
F37026\_1216 &  1.073  & 24.04 &     36.65 & S pec   &       Asymmetrical  \nl
F37029\_1427 &   0.898 & 23.67 &     36.35 &  E1/Star \nl
F37041\_1239 &  0.861 &  23.16 &     36.43 & Merger?     & Probably a protogalaxy or \nl
 & & & & &                            a collision remnant \nl
F37046\_1415 &  1.050 & 23.97 &     36.37 &  Sb:  \nl
F37055\_1129 &  1.001 &  22.37 &     36.94 & Merger? &          May contain second nucleus \nl
 & & &     & &                          within envelope  \nl
F37058\_1153 &   0.904 & 21.22 &     37.39 &  Sc  &   Prototype. See Fig.3. Two \nl
 & & &  & &                           armed spiral  \nl
F37058\_1423 &   0.970   & 22.48 &     37.10 & ?  &   Strongly nucleated \nl
F37065\_1512 & 0.840  & 22.94 &     36.48 &  E0/Star  &          Redshift refers to combined \nl
 &  &   & &        S/Ir         &         light of a and b \nl
F37078\_1605 & 0.936 &  21.88 &     37.27 & Sc   &         Clumps in disk/arms \nl
F37083\_1252 &  0.838  & 22.20 &     36.85 & Sb \nl
F37083\_1514 &  0.839   & 21.62 &     37.10 & Sb pec  & Asymmetrical bar? \nl
F37086\_1128 &  0.907  & 22.23 &     36.76 & Merger   &      At least two components,  \nl
 & & & & &                    one of these is distorted  \nl
F37089\_1202 &  0.855  & 22.90 &     36.52 & Sc     &        Single nucleus embedded \nl
 & & &   & &                               in clumpy envelope  \nl
F37096\_1055 &  0.858 &  23.16 &     36.40 & Ir   &          Ir or high SB merger \nl
F37114\_1055 &  0.855  & 22.44 &     36.73 & S pec  &     Asymmetrical  \nl
F37126\_1546 &   0.937 &  22.08 &     37.07 & E1  \nl
F37129\_1028 &  0.858  & 22.62 &     36.73 & Ir/Merger     &  Edge-on Ir, or two  \nl
 & & & & &                     component merger \nl
F37133\_1054 &  0.936 &  21.87 &     36.99 & Sa t?  \nl
F37141\_1044 &   0.821 & 22.32 &     36.70 &  E1 \nl
F37143\_1221 &  1.084  &   24.12 &     36.66 & Sa  \nl
F37154\_1212 &  1.014  & 23.25 &     37.06 & ?  &        Compact, asymmetrical \nl
F37159\_1213 &  1.020 &  23.27 &     36.86 & Sb:  &         Core embedded in extended \nl
 & & &  & &                            (tidal?) envelope  \nl
F37167\_1042 &  0.821 & 21.59 &     36.97 &  Merger &        Merger prototype,see Fig.4. \nl
 & & &  & &                          Three or four components  \nl
F37180\_1248 &  0.912 & 22.89 &     36.78 &  Sb: \nl
F37196\_1256 &  0.909  & 23.31 &     36.67 & Sa/E4  \nl
F37221\_1210 &  0.928  & 23.97 & 36.60 & Merger  &          Prototypical merger. See Fig.5. \nl
 & & &  & &            At least four components \nl
F37224\_1216 &   0.963 & 22.23 &     37.09 &  S t   & One of the components of 
F37221\_1210\tablenotemark{g} \nl
H36384\_1231 &  0.944  & 22.87 &     36.87 & S     &       Edge-on \nl
H36386\_1233 &  0.904  & 24.04 &     36.18 & Sab    &    Edge-on \nl
H36396\_1230 &  0.943  &  24.40 &     35.79 & S + E2:  &         Merger \nl
H36400\_1207 &  1.015  & 22.75 &     37.17 & Star \nl
H36408\_1203 &  1.010 & 23.49 &     36.50 & Pec/Ir &    Distorted edge-on irregular? \nl
H36408\_1205 &  0.882  & 22.94 &     36.63 & Star + Galaxy  \nl
H36431\_1242 &  0.849  &  22.34 &     37.11 & E2 \nl
H36432\_1148 &  1.010  & 23.10 &     37.01 & Sb    &      High SB two-armed spiral \nl
H36441\_1240 &  0.875  &  23.39 &     36.36 & Merger  \nl
H36443\_1133 & 1.050 &  21.96 &     37.76 & E1  \nl
H36444\_1142 &  1.020  & 24.30 &     36.68 & Merger &     Prototype. See Fig. 6. Edge-on. \nl
H36461\_1246 &  0.900  & 22.86 &     36.92 & E1 \nl
H36461\_1142 &  1.013  & 21.52 &     37.39 & Sab pec  \nl
H36463\_1404 &  0.962  & 21.69 &     37.41 & Sa pec  \nl
H36467\_1144 &  1.060  & 24.23 &     36.55 & Pec \nl
H36477\_1232 &  0.960 &  23.80 &     36.48 & E3  \nl
H36483\_1214 &   0.962  &  23.87 &     36.75 &S(B)bc t   &  Barred,or tidally \nl
 & & & & &                      distorted, spiral \nl
H36486\_1328 &  0.958 &  23.14 &     36.68 & Merger   &     Has 3 or 4 components \nl
H36490\_1221 &  0.953 & 22.59 &     36.81 & Merger?  &  High SB image has core \nl
 & & & & &                  and 4 outer lumps \nl
H36492\_1148 &  0.961 & 23.26 &     36.69 &  E4/Sa  \nl
H36493\_1155 &   0.961 &  23.36 &     36.40 & Star/E0  \nl
H36503\_1418 &  0.819  & 23.41 &     36.38 & S pec   &      Nuclear bulge embedded  \nl
 &  &  &  & &                            in eccentric ring   \nl
H36504\_1315 &  0.851 & 23.41 &     36.38 &  S pec  &   Has distorted disk \nl
H36519\_1332 &   1.087  &  23.59 &     36.76 & E1/Star  \nl
H36540\_1354 &   0.851 &  22.72 &     36.70 & S/E3 pec   &           Has distorted core  \nl
H36551\_1303 &  0.952  & 24.29 &     36.51 & E3 &    High SB \nl
H36553\_1311 &  0.968 &  22.86 &     37.11 & E/Sa: &           High SB \nl
H36555\_1353 &   1.147 &  22.85 &     36.90 & Sc/Ir: \nl
H36555\_1249 &   0.950  & 23.53 &     36.48 & Protospiral  &     See Fig.7 \nl
H36566\_1220 &   0.930  & 23.15 &     36.87 & Sa  &  High SB. Has companion  \nl
H36576\_1315 &   0.952  & 22.94 &     36.62 & S/Ir   \nl
\enddata
\tablenotetext{a}{Names are Habcde\_fghi for objects in the HDF, where the object's
J2000 coordinates are 12 ab cd.e +62 fg hi.  The initial letter is
``F'' for objects in the flanking fields.}
\tablenotetext{b}{Redshifts are from Cohen \etal\ (2000).}
\tablenotetext{c}{$R$ magnitudes are from Hogg \etal\ (2000).}
\tablenotetext{d}{Rest frame $B$ luminosities are from Cohen (2000).  
M$_B = -21.0$ (rest frame) $\equiv$ log[$L(B)$ (W)] = 36.9.}
\tablenotetext{e}{This object is a QSO from Keck spectroscopy.}
\tablenotetext{f}{These two objects have rather discrepant well
determined redshifts and are probably not physically associated.}
\tablenotetext{g}{These two objects have rather discrepant well
determined redshifts and are probably not physically associated.}
%
%
\end{deluxetable}

%
%
\begin{deluxetable}{lrrrr}
\tablenum{4}
\tablewidth{0pt}
\tablecaption{Morphological Types of Nearby and Distant 
Clusters\tablenotemark{a}}
\label{tab4}
\tablehead{\colhead{Galaxy Type} & \colhead{Shapley-Ames} & 
\colhead{HDF} & \colhead{+} & \colhead{Flanking Fields} \nl
\colhead{} &  \colhead{$z \sim 0$} & \colhead{0.25 -- 0.60} & 
\colhead{0.60-- 0.80} &  \colhead{0.80-- 1.20} \nl
\colhead{} &  \colhead{(\%)} &  \colhead{(\%)} & \colhead{(\%)} &
\colhead{(\%)} \nl
}
\startdata
  E + S0 + E/S0   &    22 &       11  &    21  &     16  \nl
  E0/Sa +S0/Sa  &    1     &    0    &        0    &          2  \nl
   Sa + Sab    &     7    &  15       & 11    &     13  \nl
  Sb +Sbc     &     27   &    26   &     23    &    10 \nl
  Sc + Sc/Ir + Scd   & 23  &     10    &          5     &         5 \nl
  Ir      &     2     &      5      &        7  &     12   \nl
  S       &       10    &      11    &      10   &     11 \nl
  ?, Pec, Protogalaxy &   7  &     16    &     19    &      12  \nl
  Merger        &    1\tablenotemark{b} &     7  &   4   &      15 \nl
\hline
Total in sample  &  936    &      50     &        70    &     
   120\tablenotemark{c}  \nl
\enddata
\tablenotetext{a}{Due to rounding errors percentages do not all add to 100.}
\tablenotetext{b}{From appendix to van den Bergh (1960c)}
\tablenotetext{c}{Includes four extragalactic objects classified 
as ``star'' and one classified as ``amorphous''}
%
%
\end{deluxetable}

\clearpage

\begin{figure}
\epsscale{0.7}
\plotone{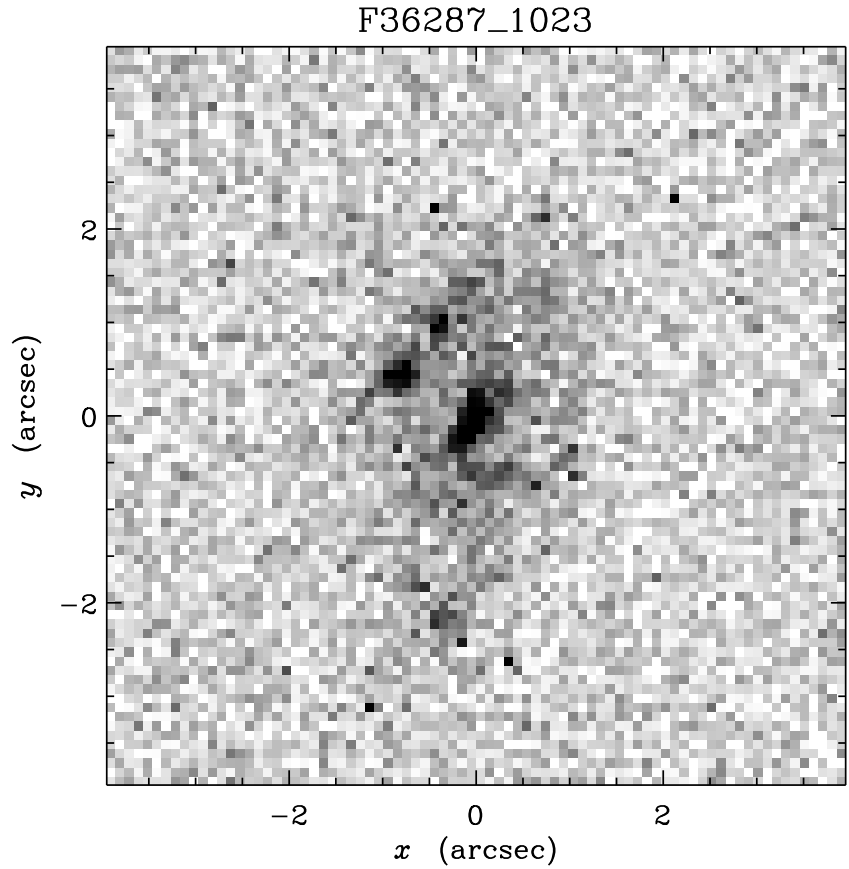}
\caption[figure1.eps] {Example at $z=0.936$ 
of merger of relatively low surface
brightness galaxies that may eventually evolve into an Sc galaxy.
In this figure, as in all the others, a 8 x 8 arcsec$^2$ section
of the WFPC2/HST image is displayed in the orientation of the
original HST image, whose scale is 0.1 arcsec/pixel.
\label{fig1}}
\end{figure}

\begin{figure}
\epsscale{0.7}
\plotone{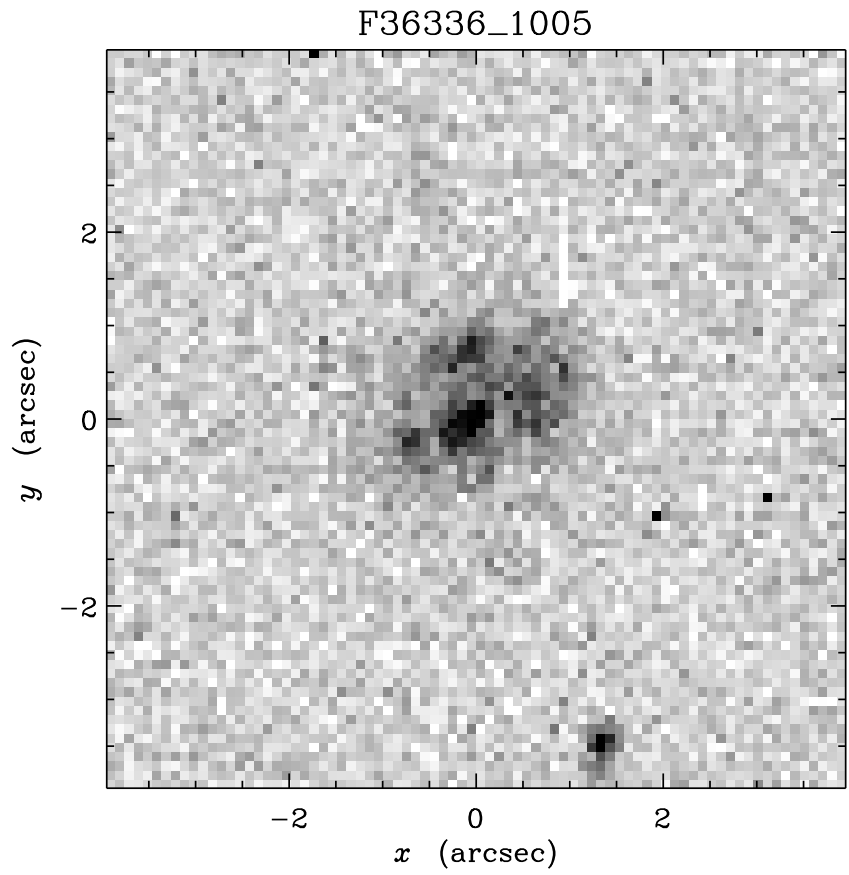}
\caption[figure2.eps] {This  $z=1.015$ object
might either be a distant Ir galaxy or a multi-component merger.
\label{fig2}}
\end{figure}

\begin{figure}
\epsscale{0.7}
\plotone{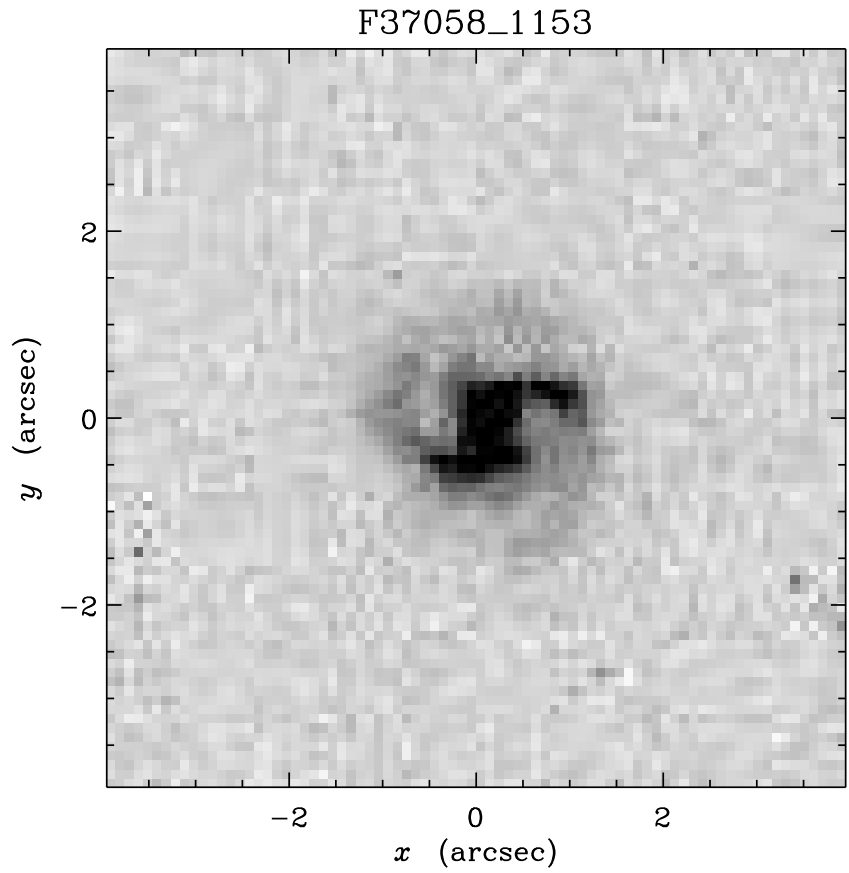}
\caption[figure3.eps] {Rare example 
of a two-armed spiral viewed at a large look-back time ($z=0.904$).
\label{fig3}}
\end{figure}

\begin{figure}
\epsscale{0.7}
\plotone{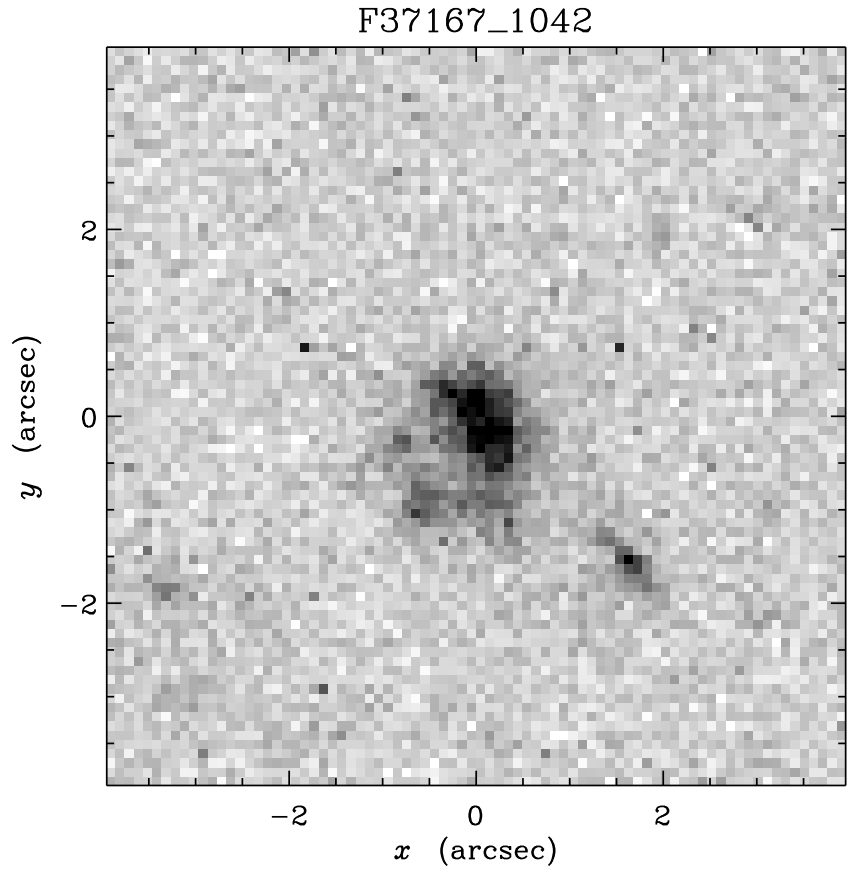}
\caption[figure4.eps] {Example (at $z=0.821$) of a
possible multi-component merger. Note 
``bits and pieces'' of other, probably more distant, objects in this
field.
\label{fig4}}
\end{figure}

\begin{figure}
\epsscale{0.7}
\plotone{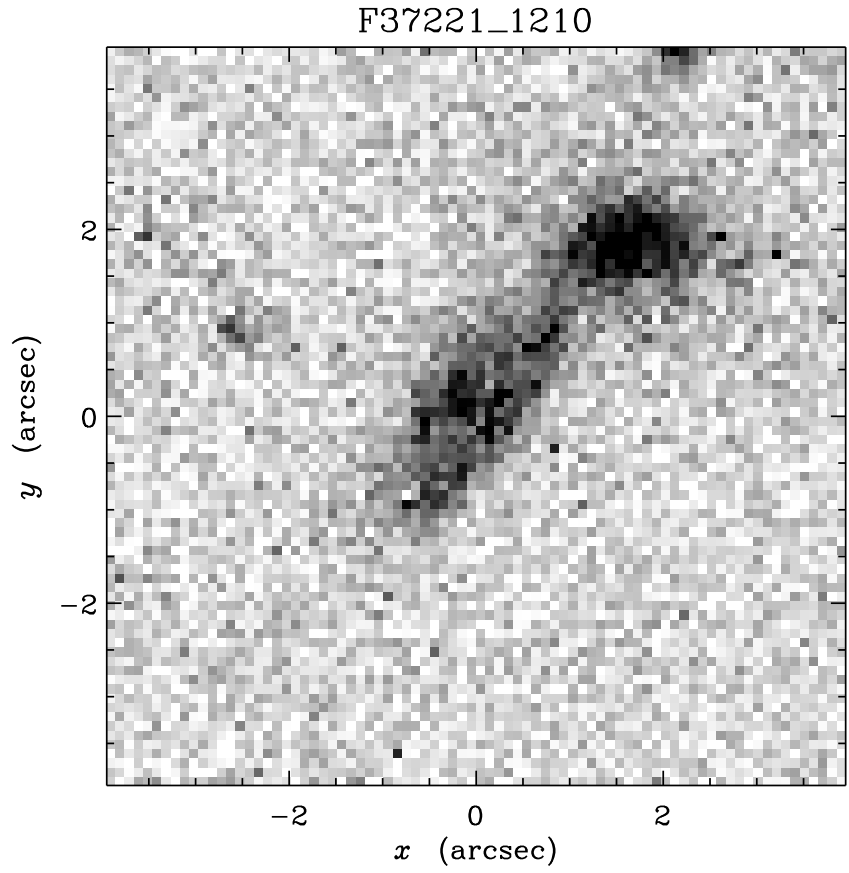}
\caption[figure5.eps] {The central part of a ``debris field'' (with $z=0.928$) 
that may eventually merge into one or more larger galaxies is shown.
\label{fig5}}
\end{figure}

\begin{figure}
\epsscale{0.7}
\plotone{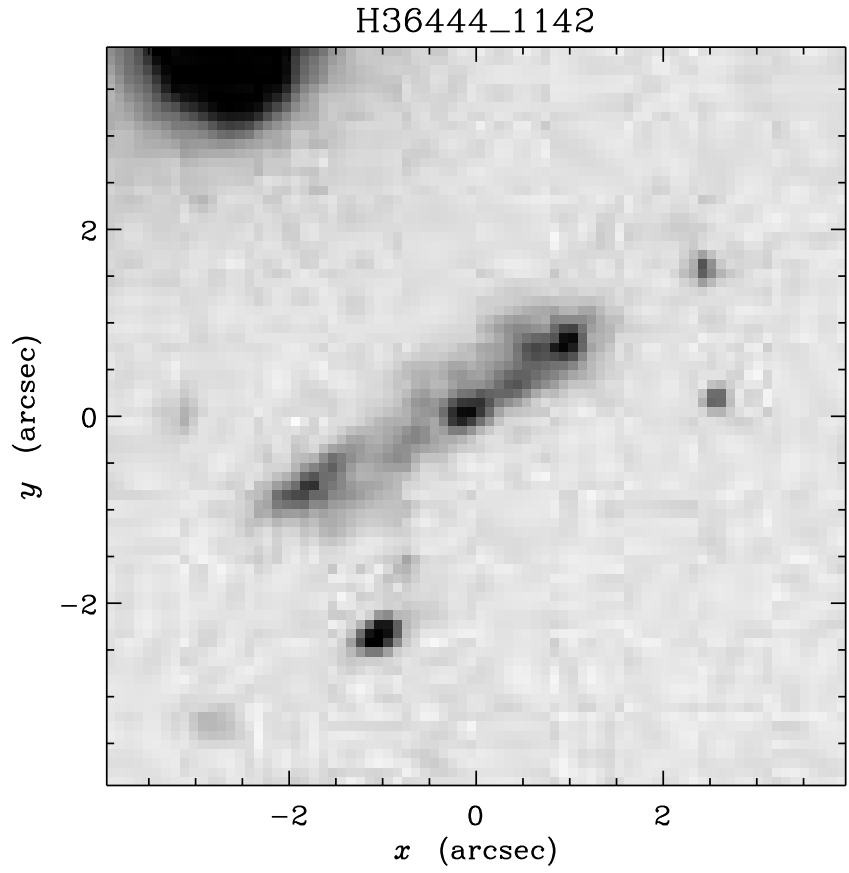}
\caption[figure6.eps] {Peculiar edge-on disk or bar-like object 
at $z=1.020$ containing
at least three merging components.
\label{fig6}}
\end{figure}

\begin{figure}
\epsscale{0.7}
\plotone{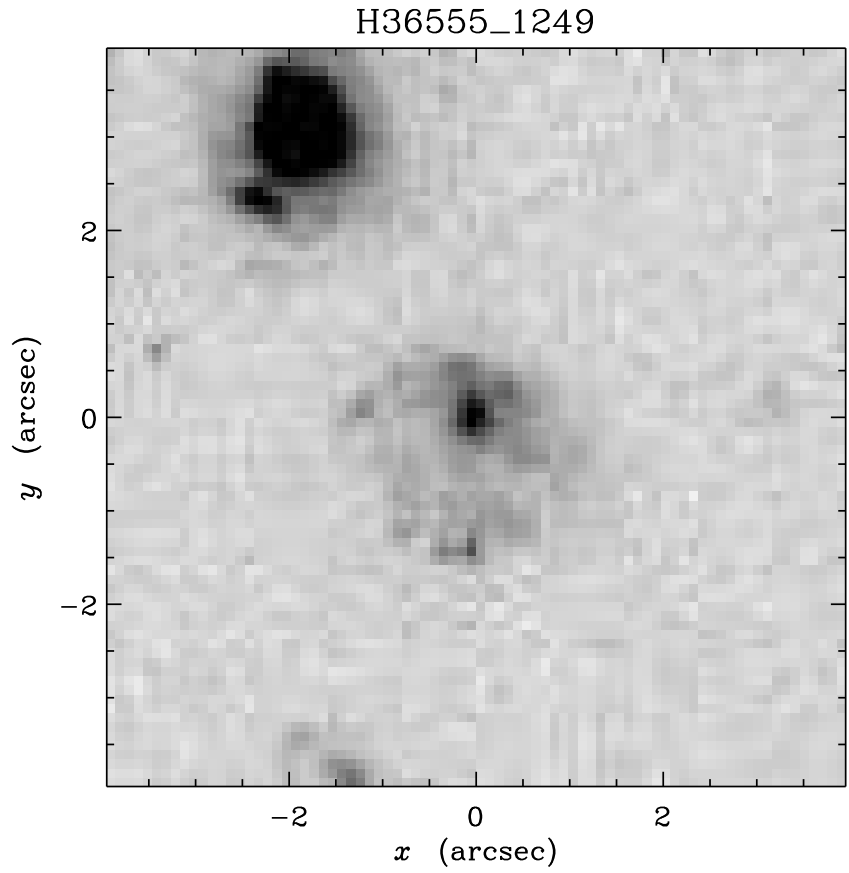}
\caption[figure7.eps] {Probable proto-spiral at $z=0.950$.  Multi-color images show that 
the dense core of this object is red (old) and that the outer
knots are blue (young).
\label{fig7}}
\end{figure}

\begin{figure}
\epsscale{0.7}
\plotone{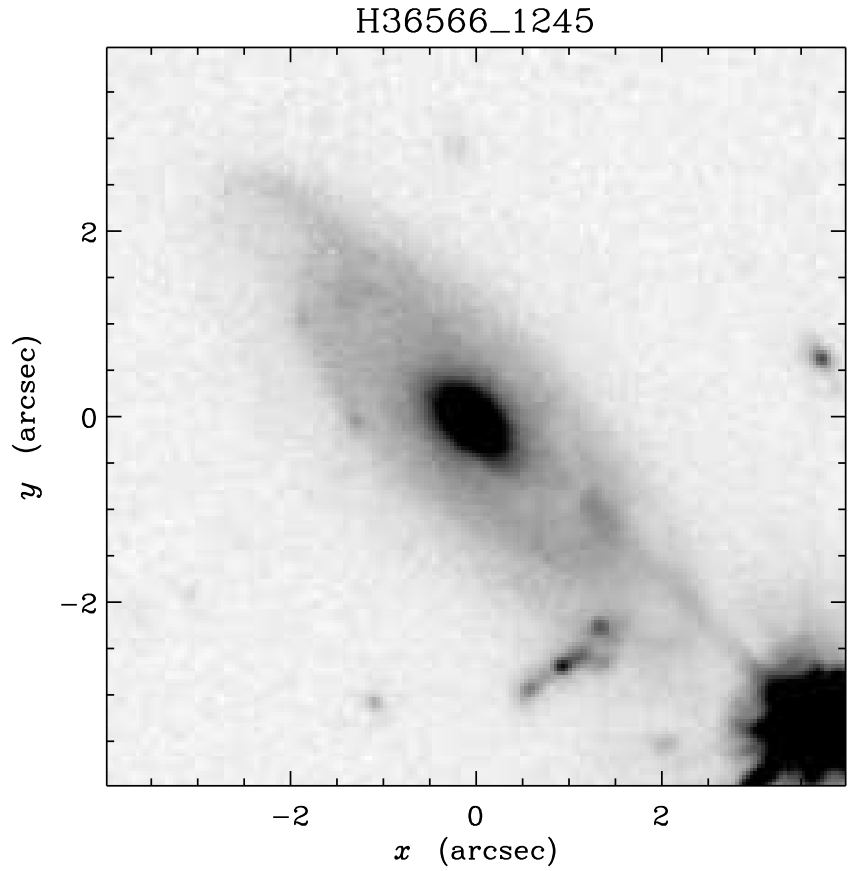}
\caption[figure8.eps] {An Sb spiral
at $z = 0.518$ which is peculiar because the spiral arms in the 
disk are underdeveloped compared to typical objects of similar 
type at $z\sim0.0$. Note the multi-component (background) merger below
and to the right of this object.
\label{fig8}}
\end{figure}

\begin{figure}
\epsscale{0.7}
\plotone{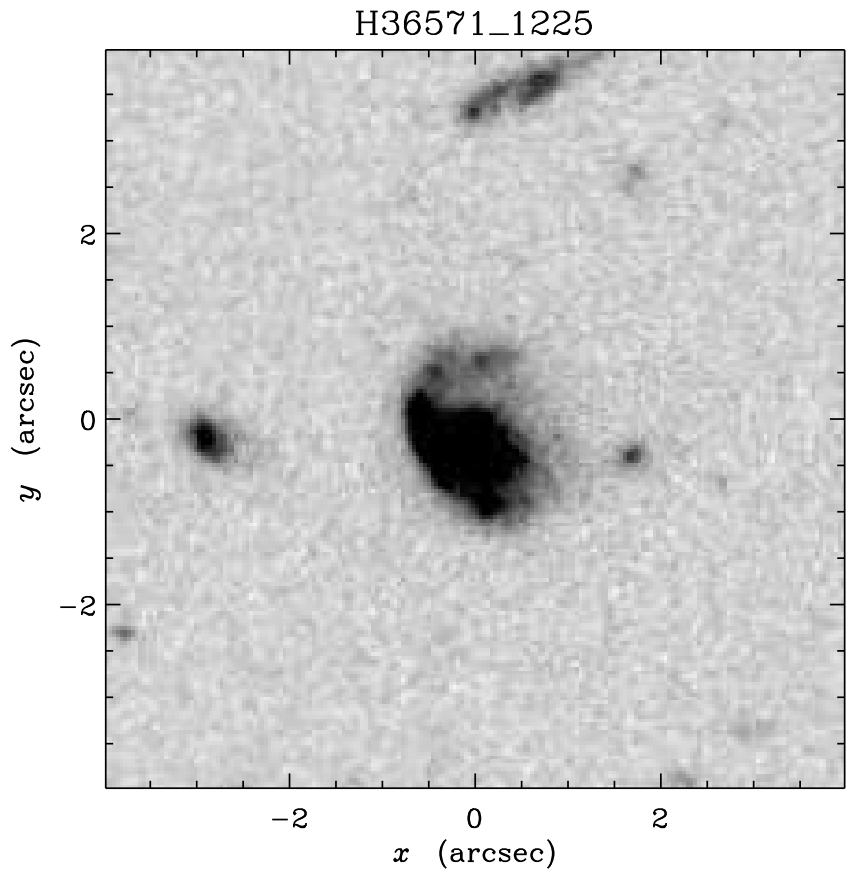}
\caption[figure9.eps] {Note the malformed single
spiral arm of this peculiar Sc galaxy at $z = 0.561$.
\label{fig9}}
\end{figure}

\begin{figure}
\epsscale{0.7}
\plotone{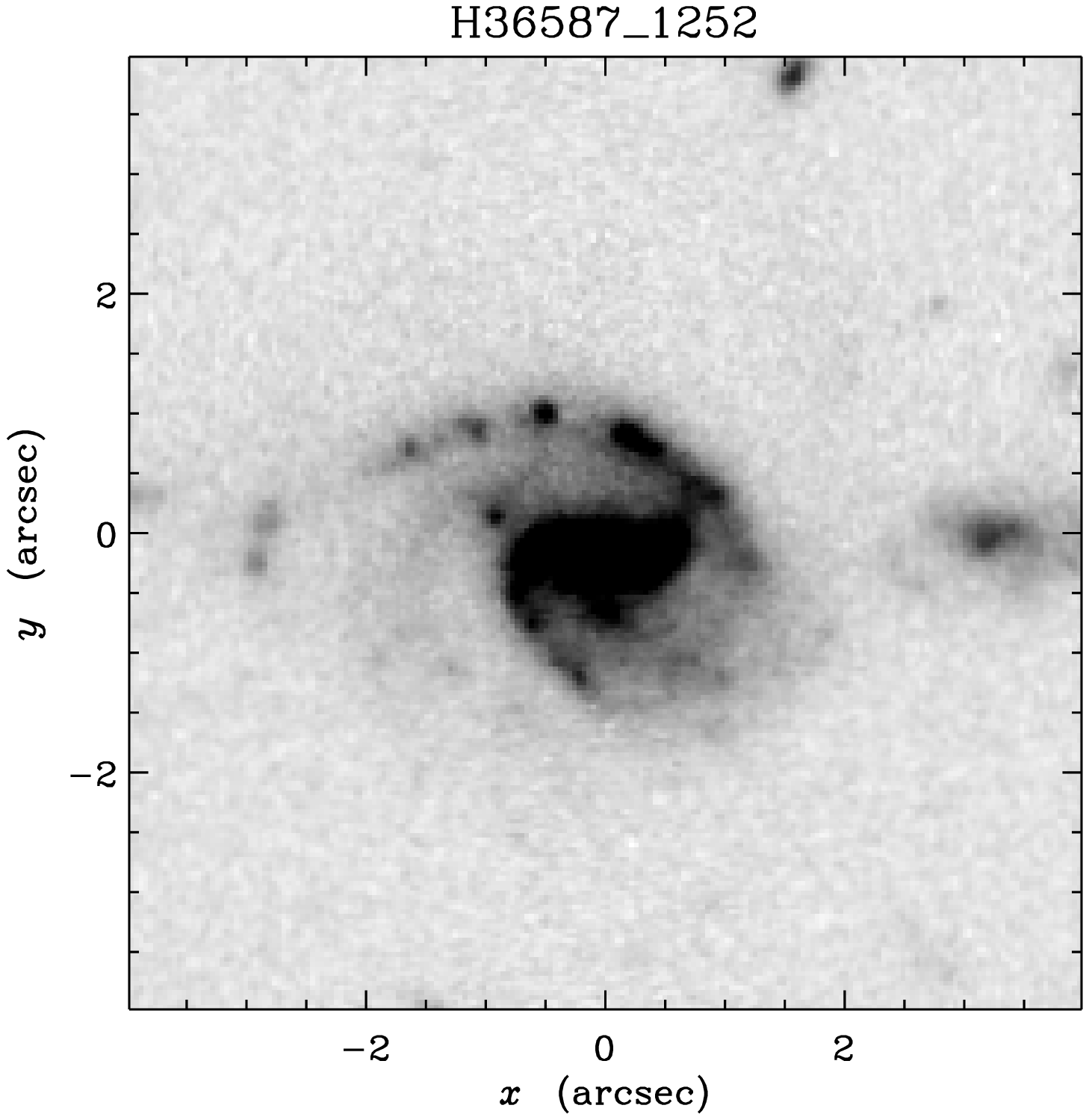}
\caption[figure10.eps] {The most distant ``certain'' barred spiral in the present sample (at $z=0.321$) is shown.
Note the difference in the lengths of  spiral arms in this SBc pec 
galaxy
\label{fig10}}
\end{figure}


\clearpage

\begin{references}

\reference{} Abadi, M.~G., Moore, B. \& Bower, R.~G., 1999, \mnras, 308, 947

\reference{} Abraham, R. G. 1999, in {\it{Interactions of Galaxies at 
Low and High Redshifts}},
                Eds. J. E. Barnes and D. B. Sanders (Dordrecht:Kluwer), p.11

\reference{} Abraham, R. G., 2000, {\it {Toward a New Millennium in Galaxy Morphology}}, eds. D. L. Block \etal\ (Dordrecht: Kluwer) in press,
(astro-ph/0006166)

\reference{} Abraham, R. G., Tanvir, N. R., Santiago, B. X., Ellis, R. S.,
               Glazebrook, K. \& van den Bergh, S., 1996b, \mnras, 279, L47

\reference{} Abraham, R. G., Ellis, R. S., Fabian, A. C., Tanvir, N. \& Glazebrook, K. 1999a \mnras, 303, 641

\reference{} Abraham, R. G., Merrifield, M. R., Ellis, R. S., Tanvir, N., Brinchmann, J. 1999b, \mnras, 308, 569

\reference{} Abraham, R. G., van den Bergh, S., Glazebrook, K.,
Ellis, R. S., Santiago, B.X., Surma, P. \& Griffiths, R.E., 1996b,
\apjs, 107, 1

\reference{} Ames, M., 1997, The (London) Times, 1997 August 21

\reference{} Bell, E. F. \& de Jong R. S., 2000, \mnras, in press

\reference{} Boissier, S. \& Prantzos, N., 2000, \mnras, 312, 398

\reference{} Brinchmann, J. \etal\, 1998, \apj, 499, 112

\reference{} Bunker, A., Spinrad, H., Stern, D., Thompson, R., Moustaka, L.,
Davis, M. \& Dey, A., 2000, to appear in {\it{Galaxies in the Young
Universe}}, ed. H. Hippelein \& K. Meisenheimer, Astro-ph/0004348  

\reference{} Butcher, H. \& Oemler, G., 1978, \apj, 219,18

\reference{} Carlberg, R. G . \etal\, 2000, \apj, 532, L1


\reference{} Cohen, J. G., 2000, \aj\ (submitted)

\reference{}  Cohen, J.G., Cowie, L.L., Hogg, D.W., Songaila, A.,
Blandford, R., Hu, E.M. \& Shopbell, P., 1996, \apjl, 471, L5

\reference{} Cohen, J. G., Hogg, D. W., Blandford, R. D., Cowie, L. L., Hu, E., 
Songaila, A., Shopbell, P. \& Richberg, K., 2000, \apj, 538, 29

\reference{} Corbin, M. R., Vacca, W. D., O'Neil, E., Thompson, R. I.,
Rieke, M. J., \&
              Schneider, G., 2000, \aj, 119, 1062

\reference{} Cowie, L. L., Hu, E. M. \& Songaila, A., 1995, \aj, 110, 1576

\reference{} Dressler, A., 1980, \apj, 236, 351

\reference{} Driver, S. P., Fernandez-Soto, A., Couch, W. J., Odewahn, S. C.,
   Windhorst, R. A., Phillipps, S., Lanzetta, K., \& Yahil, A.,
   1998, \apj, 496,L93

\reference{} Eskridge, P. B. \etal\, 2000, \aj, 119, 536

\reference{} Falomo, R., Urry, C. M., Pesce, J. E., Scapa, R.,
Giavalisco, M. \& Treves, A., 
              1997, \apj, 476, 113

\reference{} Fasano, G., Poggianti, B. M., Couch, W. J., Bettoni, D.,
Kjaergaard, P. \& Moles, M., 2000, \apj\ (in press) (astro-ph/0005171)  

\reference{} Ferguson, H. C., Dickinson, M. \& Williams, R., 
2000, ARAA (in press)


\reference{} Graham, A.~W. \& Prieto, M., 1999, \apjl, 524, L23

\reference{} Gunn, J. E. \& Gott, J. R., 1972, \apj, 176, 1

\reference{} Hartley, L. P., 1953 {\it{The Go-Between}}, (London: H.Hamilton)

\reference{}  Hogg D.~W., Pahre M.~A., Adelberger K.~L., Blandford R., 
Cohen J.~G.,
Gautier T.~N., Jarrett T., Neugebauer G. \& Steidel C.~C., 1999, \apjs,
127, 1

\reference{} Hubble, E. 1936, {\it{The Realm of the Nebulae}}, (New Haven: Yale University Press), p.45

\reference{} Le F\'evre,O. \etal\, 2000, \mnras, 311, 565

\reference{} Lilly, S. \etal\, 1998, \apj, 500, 75

\reference{} Marleau, F. R. \& Simard, L., 1998, \apj, 507, 585 

\reference{} Morgan, W. W., 1958, \pasp, 70, 364

\reference{} Morgan, W. W., 1959, \pasp, 71, 394

\reference{} Odewahn, S. C., Windhorst, R. A., Driver, S. P. \& Keel, W. C.,1996,
\apj, 472, L13

\reference{} Pfenniger, D.,  Martinet, L. \& Combes, F., 1996, in 
{\it{New Extragalactic Perspectives in South Africa}}, ed.
D.L.Block \& J.M.Greenberg, (Dordrecht: Kluwer),p.291

\reference{} Quilis, V., Moore, B. \& Bower, R., 2000, Science, 288, 1617

\reference{} Raha, N., Sellwood, J.A., James, R.A. \& Kahn, F.D., 1991,
Nature, 352, 411

\reference{} Sandage, A. \& Tammann, G. A., 1981,
{\it{A Revised Shapley-Ames Catalog of Bright
              Galaxies}}, (Washington: Carnegie Institution)

\reference{} Schade, D., Lilly, S.J., Crampton, D., Ellis, R.S.,
Le F\'evre, O., Hammer, F., Brinchmann, J., Abraham, R., Colless, M.,
Glazebrook, K., Tresse, L. \& Broadhurst, T., 1999, \apj, 525, 31

\reference{} Simard, L., Koo. D.C., Faber, S.M., Sarjedini, V.L.,
Vogt, N.P., Phillips, A.C., Gebhardt, K., Illingworth, G.D. \&
Wu, K.L., 1999, \apj, 519, 563

\reference{} Toomre, A., 1977, in {\it{The Evolution of Steller Populations}}, Eds. B. M. Tinsley
              \& R. B. Larson, 1977 (New Haven: Yale Observatory), p.420

\reference{} van den Bergh, S., 1960a, \apj, 131, 215

\reference{} van den Bergh, S., 1960b, \apj, 131, 558

\reference{} van den Bergh, S., 1960c, Pub.David Dunlap Obs., 2, 159

\reference{} van den Bergh, S., 1989, \aj, 97, 1556

\reference{} van den Bergh, S., 1998, {\it{Galaxy Morphology and 
Classification}}, (Cambridge: Cambridge Univ. Press)

\reference{} van den Bergh, S., Abraham, R. G., Ellis, R. S., Tanvir, N. R.,
   Santiago, B. X. \& Glazebrook, K. G., 1996, \aj, 112, 359

\reference{} van den Bergh, S. \& Pierce, M. J. 1990, \apj, 364, 444

\reference{} Visvanathan, N. \& van den Bergh, S., 1992, \aj, 103, 1057

\reference{} Williams, R. E. \etal\, 1996, \aj, 112, 1335

\end{references}
\end{document}